\documentclass{osa-article}

\journal{osajournal}

\articletype{Research Article}

\begin{document}

\title{Photochromism in a Hexagonal Boron Nitride Single Photon Emitter}

\author{Matthew A. Feldman,\authormark{1,2,*} Claire E. Marvinney,\authormark{3} Alexander A. Puretzky,\authormark{4}, Benjamin J. Lawrie,\authormark{5,**}}

\address{\authormark{1}Department of Physics and Astronomy, Vanderbilt University, Nashville TN 37235\\
\authormark{2}Computational Sciences and Engineering Division, Oak Ridge National Laboratory, Oak Ridge TN 37831\\
\authormark{3}Physics Division, Oak Ridge National Laboratory, Oak Ridge TN 37831\\
\authormark{4}Center for Nanophase Materials Sciences, Oak Ridge National Laboratory, Oak Ridge TN 37831\\
\authormark{3}Physics Division, Oak Ridge National Laboratory, Oak Ridge TN 37831}
\email{\authormark{*} matthew.feldman@vanderbilt.edu, \authormark{**} lawriebj@ornl.gov}



\date{\today}




\begin{abstract}
  Solid-state single-photon emitters (SPEs) such as the bright, stable, room-temperature defects within hexagonal boron nitride (hBN) are of increasing interest for quantum information science applications. To date, the atomic and electronic origins of SPEs within hBN are not well understood, and no studies have reported photochromism or explored cross-correlations between hBN SPEs. Here, we combine irradiation-time dependent measures of quantum efficiency and microphotoluminescence ($\mu$PL) spectroscopy with two-color Hanbury Brown-Twiss interferometry to enable an investigation of the electronic structure of hBN defects. We identify photochromism in a hBN SPE that exhibits cross-correlations and correlated quantum efficiencies between the emission of its two zero-phonon lines.    
\end{abstract}

\section{INTRODUCTION}
Solid-state single-photon emitters (SPEs) are of increasing interest as a source of non-classical light for quantum computation, quantum communication, and quantum sensing applications \cite{aharonovich2016solid,awschalom2013quantum, awschalom2018quantum}.  Defects in hexagonal boron nitride (hBN) have emerged as notable SPEs due to their bright, stable, room-temperature emission across the 
visible spectrum \cite{tran2016quantum}. Recent advances in the characterization and control of hBN SPEs include the characterization of spin states in a defect ensemble with optical detected magnetic resonance \cite{gottscholl2020initialization} that could be used for quantum memories\cite{atature2018material}. They also include charge state initialization \cite{khatri2020optical} that could enable coherent optical control applications \cite{konthasinghe2019rabi}.  Furthermore, strain localization \cite{proscia2018near} and strain tuning \cite{grosso2017tunable,mendelson2019strain} could enable the design of deterministic indistinguishable single photon sources.

Despite these advances in state preparation, readout, and process control, and despite substantial theoretical~\cite{Twafik2017,Abdi2018,PhysRevB.97.064101,Grosso2020} and microscopic~\cite{hayee2020revealing} analysis, the atomic origins and electronic structure of hBN SPEs is still poorly understood. To date, defects have been categorized phenomenologically. Initial reports identified group I and group II hBN SPEs based on the difference in their electron-phonon coupling \cite{tran2016robust}. More recent research has demonstrated the existence of  four species of hBN emitters spanning the visible spectrum with correlated microphotoluminescence ($\mu$PL), cathodoluminescence, and nanoscale strain mapping \cite{hayee2020revealing}. Further research has identified photochemical effects such as bleaching of hBN emitters under 405~nm excitation\cite{shotan2016photoinduced}, or activation of emitters with electron-beam irradiation \cite{ngoc2018effects}. To date polarimetry studies of hBN emitters under i) 472 nm and 532 nm excitation and ii) tunable strain have exhibited a misalignment in their absorption and emission dipole moments supporting the claim of a third excited bright state in hBN defects or complexes\cite{acs.nanolett.6b01987,PhysRevLett.119.057401,adma.201908316}. However, no studies to date have directly reported photochromism in hBN SPEs or explored cross-correlations between hBN SPEs or bright state transitions of a hBN SPE.

\begin{figure*}
    \centering
    \includegraphics[width=\textwidth]{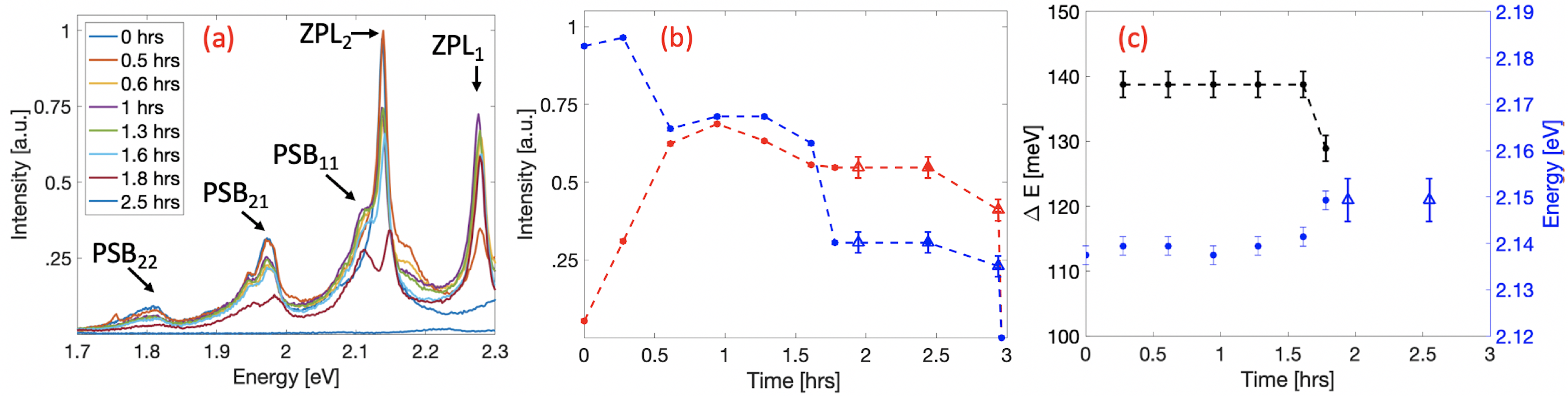}
    \caption{Laser irradiation-dependent spectroscopy of a single defect pumped with a cw 405 nm laser. (a) The $\mu$PL spectra of ZPL transitions ZPL$_1$ and ZPL$_2$. Here ZPL$_1$ has a one-phonon sideband (PSB$_{11}$) 166~meV redshifted from  ZPL$_1$, and ZPL$_2$ has one-phonon (PSB$_{21}$) and two-phonon sidebands (PSB$_{22}$) 166 and 326~meV, respectively, redshifted from ZPL$_2$. (b) The relative quantum efficiencies of ZPL$_1$ (red) and ZPL$_2$ (blue) (normalized to the peak ZPL$_2$ efficiency) show enhancement and partial quenching within the first half hour of irradiation, respectively. For the following hour, the quantum efficiencies remain  stable, after which ZPL$_2$ undergoes a second partial quenching. The ZPL$_1$ and ZPL$_2$ quantum efficiencies remain stable for an hour prior to their simultaneously quenching. (c) The energy difference (black) between ZPL$_1$ and ZPL$_2$ remains constant until the second partial quenching in ZPL$_2$ occurs leading to a 10~meV spectral jump in the energy of ZPL$_2$ (blue). Triangles indicate measurements made using filtered singles counts.}
    \label{fig:fig1}
\end{figure*}

Here we use $\mu$PL spectroscopy to study the photostability of defects in few-layer hBN flakes in air when optically pumped with greater photon energy than the activation energy for the photochemical decomposition of hBN. Further, we characterize the cross-correlations between zero-phonon lines (ZPLs) that exhibit correlated quantum efficiencies with spectrally resolved two-color Hanbury Brown-Twiss (HBT) interferometry.  While we have previously used two-color HBT interferometry with photo-stable emitters in hBN under vacuum to verify that the broad emission bands redshifted 166$\pm$0.5~meV and 326$\pm$0.5~meV from the ZPL are optical one- and two-phonon sidebands respectively\cite{feldman2019phonon}, the cross-correlated ZPLs studied in this work have separation energies 20~meV below the known optical phonon modes of hBN \cite{vuong2016phonon,khatri2019phonon} and localized phonon resonances fail to explain the observed spectrum\cite{Grosso2020}. Combining irradiation-time dependent measures of quantum efficiency and $\mu$PL spectroscopy with two-color HBT interferometry enables this new investigation of the electronic structure of hBN defects.
\section{Results and Discussion}
 \subsection{Microphotoluminescence Results} We investigate defects in three-to-five layer hBN using the same sample from a previous study\cite{feldman2019phonon}. Microphotoluminescence spectroscopic data was collected for each defect using a custom-built room-temperature confocal microscope.  Eleven defects with ZPLs ranging from 2.15 to 2.9~eV were observed. The majority of defects measured were identified as group I emitters with $\sim$10~meV linewidth ZPLs and one-phonon doublets and two-phonon sidebands 166 and 326 meV redshifted from the ZPL\cite{tran2016robust,feldman2019phonon}. Out of these, seven defects photobleached consistent with previous $\mu$PL studies of defects under oxygen rich environments
 \cite{shotan2016photoinduced}. Among the defects that photobleached, two ZPLs within a diffraction-limited confocal volume demonstrated a correlated enhancement and quenching in their quantum efficiency under 405~nm excitation. Figure 1a shows the $\mu$PL spectra measured for that site. ZPL$_1$ (2.28 eV) has a one-phonon (PSB$_{11}$) sideband 166$\pm$2~meV red-shifted with respect to ZPL$_1$, and  ZPL$_2$ (2.14 eV) has one-phonon (PSB$_{21}$) and two-phonon (PSB$_{22}$) sidebands 166$\pm$2 and 326$\pm$2~meV redshifted from ZPL$_2$. In the first half-hour of irradiation, the quantum efficiency of ZPL$_1$ and PSB$_{11}$ increases while ZPL$_2$, PSB$_{21}$ and PSB$_{22}$ decrease as seen in Figure 1a-b. The quantum efficiencies for ZPL$_1$ and ZPL$_2$ and their corresponding phonon sidebands then equilibriate for an hour until ZPL$_2$ is again partially quenched. Prior to this second quench in ZPL$_2$ the energy difference between ZPL$_1$ and ZPL$_2$ was 139$\pm$2~meV; afterwards the energy difference decreases to 129$\pm$2~meV due to a 10~meV blue shift in ZPL$_2$ as seen in Figure 1c. The blueshifted ZPL$_2$ and the ZPL$_1$ energies were found not to undergo any other spectral jumps and were stable to within the spectral resolution of the $\mu$PL measurements before both ZPL$_1$ and ZPL$_2$ simultaneously quenched as seen in Figure 1b 
 and Figure S1 in the Supporting Information.

\subsection{Microphotoluminescence Discussion} These data support the claims that i) ZPL$_1$ and ZPL$_2$ are correlated with one another and are excited-state transitions of the same defect or complex and ii) changes in the local crystal symmetry, the charge state of the defect, or the mobility of the defect are caused by a photochemical reaction. The stable blueshift in ZPL$_2$ and its simultaneous reduction in quantum efficiency suggests that the local strain environment was modified as previously reported in first-principle studies of carrier recombination dynamics in hBN defects\cite{PhysRevB.100.081407}. Previous work has shown that the quantum efficiency of defects in one-to-five layer hBN flakes are enhanced under 405~nm excitation whereas only a fraction of defects fluoresce under 532~nm excitation\cite{shotan2016photoinduced}. The difference in absorption of 405~nm versus 532~nm light has been attributed to the presence of additional excited states that compete with lower energy transitions. The same study demonstrated photobleaching of defects under 405~nm excitation whereas defects were photostable under 532~nm excitation with the sample in ambient conditions. These observations are in agreement with this work and previous work that has shown a reduction in fluorescence and photobleaching of visible and ultraviolet transitions for near surface defects pumped at energies exceeding the activation energy for the photochemical decomposition of hBN of 2.57 eV\cite{Kanaev2004} under 99$\%$ molecular oxygen or air while these defects' $\mu$PL are photostable under vacuum~\cite{Museura2007,Berzina2016,feldman2019phonon}. Recent correlated cathodoluminescence and $\mu$PL studies also identified a class of hBN emitters comparable to those shown in Fig. 1a that emit at 2.15-2.3~eV, are formed in strained environments,and are stable when pumped with 80~keV irradiation while under vacuum~\cite{hayee2020revealing}. The same report suggested that the emitters may be complexes involving oxygen based defects such as O$_{2B}$V$_N$ \cite{hayee2020revealing}. Alternatively, hBN ZPLs at 2.15-2.2 eV with misaligned emission and absorption dipole moments have shown exceedingly large tunability ($\sim$10 nm) under variable strain fields\cite{adma.201908316}. This suggests that these defects are not inversion symmetric, as the ZPLs of inversion symmetric quantum emitters are typically only tunable in the $\sim$10-100 GHz range\cite{PhysRevB.97.205444}. For hBN emitters which are not inversion symmetry and are in the visible range, N$_B$V$_N$, C$_B$V$_N$ are viable defect candidates\cite{adma.201908316}. Our findings provide a strong indication of photochemical reactions of near-surface defects or hBN lattice atoms with O$_2$. The defects remained dark after a month, suggesting that the defect state is either destroyed by the photochemical process or pumped into an very long-lived dark state. 

\begin{figure*}[t!]
    \centering
    \includegraphics[width=\textwidth]{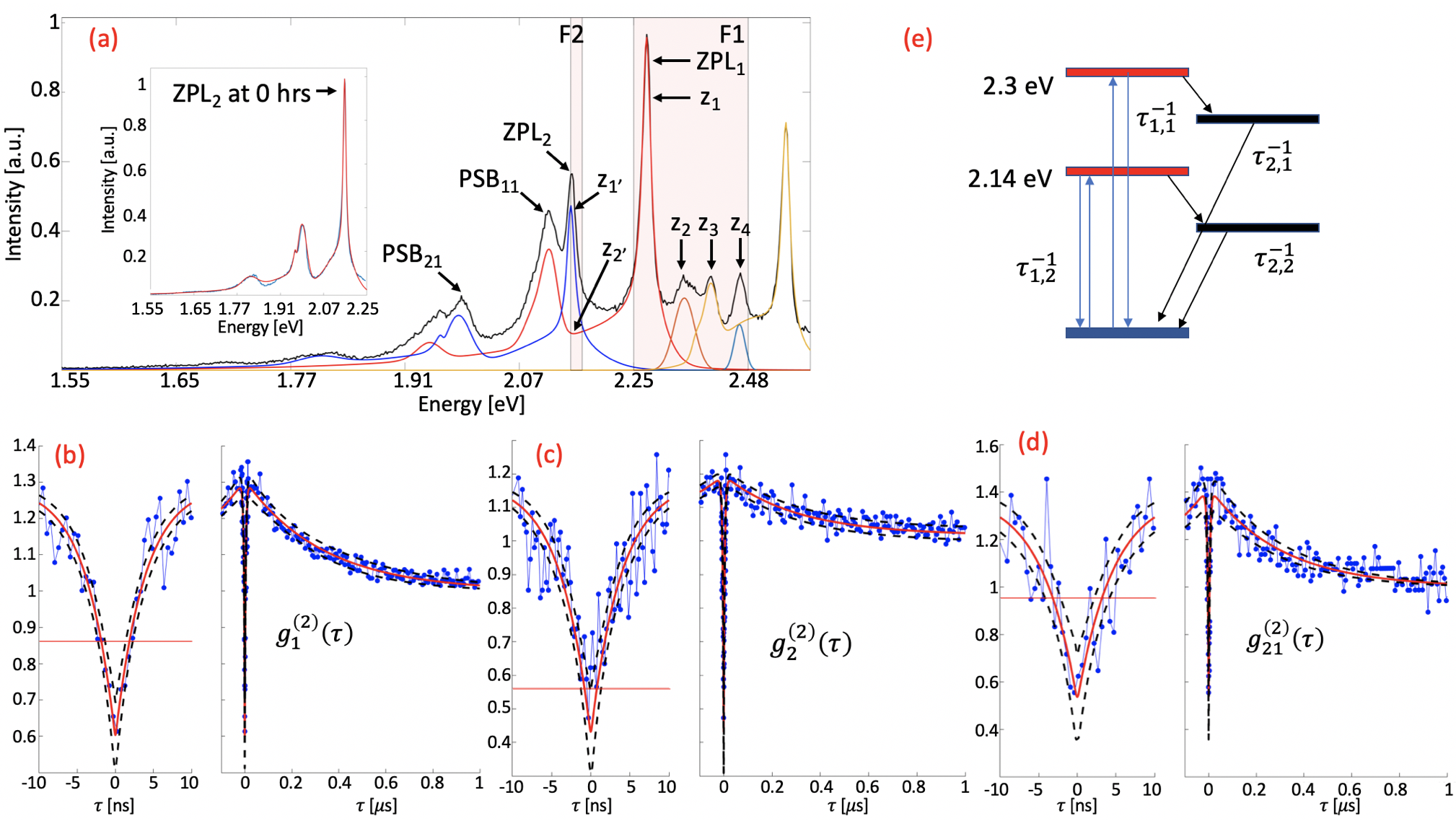}
    \caption{Two-color Hanbury Brown-Twiss interferometry. (a) Spectral lineshapes each comprising a ZPL and their corresponding phonon sidebands for ZPL$_1$ and ZPL$_2$ after ZPL$_2$ undergoes a spectral jump with filters F1 and F2 indicating spectral-bands passed to the HBT interferometer's arms. Fits to the ZPL$_1$ (red), ZPL$_2$ (blue) lineshapes and uncorrelated emitters are used to estimate the probability (z$_i$,z$_j^{'}$) that a transition contributes to the $\mu$PL (black) collected in each arm. The lineshapes overlapping with F1 and F2 are labeled by their probability of contributing to the total intensity in each arm. The inset shows the best fit for ZPL$_2$ at ~0 hrs. (b) The auto-correlations for ZPL$_1$ and (c) ZPL$_2$ are antibunched and (d) the cross-correlations between ZPL$_1$ and ZPL$_2$ are anticorrelated. The distance of $g^{(2)}_i(0)$ from the threshold for single-photon-emission (indicated by the red horizontal line) exceeds five standard deviations, $\sigma$. Here the red-dashed lines are the 5$\sigma$ bounds for  $g^{(2)}_i(\tau)$. A proposed energy diagram for the suspected defect complex is presented in (e).}
    \label{fig:example}
\end{figure*}

\subsection{Two-Color HBT Interferometry Results} To test the hypothesis that ZPL$_1$ and ZPL$_2$ are excited-state transitions of the same defect, we employed two-color HBT interferometry on ZPL$_1$ and ZPL$_2$ after ZPL$_2$ spectrally jumped 10~meV. Filters F1 and F2 (illustrated in Figure 2a) selected the luminescence of ZPL$_1$ and ZPL$_2$ in each arm of the interferometer. Because of the presence of additional defects in the spectrum shown in Fig. 2a, the background counts in each channel could not be attributed solely to a Poissonian background. Instead, the background is modeled as defect emission that is uncorrelated with ZPL$_1$ and ZPL$_2$. To determine the contribution of each emitter to the counts in each channel, fits of the lineshapes corresponding to ZPL$_1$ and ZPL$_2$ and their respective phonon sidebands were made as shown in Figure 2a. Here ZPL$_1$ and ZPL$_2$ were fit using a sum of a Lorentzian and Gaussian distributions centered at their ZPL and phonon-sideband energies, respectively, as described in the Supplemental Material. The lineshape of the ZPL$_2$ emission was assumed to be fixed irrespective of spectral jumps and so only the amplitude and peak energy parameters for ZPL$_2$ were left free for all irradiation times. All peaks attributed to uncorrelated background emission were fit using the same lineshape function or with a Gaussian distribution. 

The probability $z_i$ and $z_{i'}$ that each emitter will contribute to the counts on arm one and two, respectively, of the HBT interferometer is the overlap of the filter transfer function with the total emission of the $i^{th}$ emitter divided by the total emission of all emitters in the filter-band~\cite{bommer2019new}.  The measured coincidence counts corresponding to the auto-correlations for ZPL$_1$ and ZPL$_2$ as well as the cross-correlation between the ZPL$_1$ and ZPL$_2$ were calculated with this assumption in mind. The bunching observed in the coincidence counts for $|\tau| > 0$ for both transitions indicates a shelving state is present in each ZPL and we interpret these data assuming a three-level model. The auto-correlations for ZPL$_1$ and ZPL$_2$ and the cross-correlations between each are given by
\begin{align}
    g^{(2)}_{1}(\tau) &= (z^2_{1}+z^2_{2}+z^2_{3}+z^2_{4})g^{(2)}_{\rho1}(\tau) +2(z_1z_2+z_1z_3+z_1z_4+z_2z_3+z_2z_4+z_3z_4),  \\  
    g^{(2)}_2(\tau)&= (z^2_{1'}+z^2_{2'})g^{(2)}_{\rho2}(\tau) 
    + z_{1'}z_{2'}(g^{(2)}_{21}(\tau)+g^{(2)}_{12}(\tau)), \\
    g^{(2)}_{21}(\tau) &= z_1z_{1'}g^{(2)}_{\rho21}(\tau) + z_1z_{2'}g^{(2)}_{1}(\tau)
    +z_2+z_3+z_4, \\
    g^{(2)}_{\rho i}(\tau) &= 1-\rho_i^2[(1+a_i)e^{-|x-x_{oi}|/\tau_{1i}}
    -{a_i}e^{-|x-x_{oi}|/\tau_{2i}}], 
\end{align}

where  $g^{(2)}_{1}(\tau)$ and $g^{(2)}_{2}(\tau)$ are the auto-correlation functions for ZPL$_1$ and ZPL$_2$, $g^{(2)}_{21}(\tau)$ is the cross-correlation between the zero-phonon lines, $g^{(2)}_{\rho i}(\tau)$ is the three-level model for each correlation function with Poisson-background contribution $\rho_i$, shelving parameter $a_i$, excited state lifetime $\tau_{1i}$, and shelving state lifetime $\tau_{2i}$. A full derivation of the auto- and cross-correlations can be found in the Supporting Information. The threshold for single-photon emission for auto- and cross-correlations is given by $g^{(2)}_{limit,i} = \frac{1}{2}(1+\rho_i^2a_i)$.

The cross-terms for the $g^{(2)}_{1}(\tau)$, $g^{(2)}_{2}(\tau)$ and $g^{(2)}_{21}(\tau)$ degrade the single-photon purity of the emitter while the like-terms dampen the expected shelving amplitude. Figure 2b-d show the auto- and cross-correlations as well as best-fit and corresponding 5$\sigma$ confidence intervals for $g^{(2)}_{1}(\tau)$, $g^{(2)}_{2}(\tau)$ and $g^{(2)}_{21}(\tau)$, respectively. It is clear that the auto- and cross-correlations confirm single-photon emission and photochromism between ZPL$_1$ and ZPL$_2$ as the fit for $g^{(2)}_i(0)$ is at least five standard deviations from the limit for single-photon emission. The overlap of ZPL$_1$ and ZPL$_2$ leads to additional time dependent cross-terms terms in the auto-correlation for ZPL$_2$ ($g^{(2)}_{12}(\tau)$, $g^{(2)}_{21}(\tau)$) and the cross-correlation between ZPL$_1$ and ZPL$_2$ ($g^{(2)}_{1}(\tau)$) but their contribution to the anti-bunching and anti-correlations are found to be $\sim$10$\%$. Furthermore all fits for $g^{(2)}_1(\tau)$, $g^{(2)}_2(\tau)$ and $g^{(2)}_{21}(\tau)$ include cross-terms and thus the background-free cross-correlation between ZPL$_1$ and ZPL$_2$ and the respective anti-bunching for each zero-phonon line would increase the distance $[g^{(2)}_{limit,i}-g^{(2)}_i(0)]$. 

Additionally, while the defect photobleached prior to collecting $g^{(2)}_{12}(\tau)$, the large cross-correlation and single-photon purity remained consistent with $[g^{(2)}_{limit,i}-g^{(2)}_i(0)]> 5\sigma$ under the assumptions that $g^{(2)}_{12}(\tau) \approx g^{(2)}_{21}(\tau)$, $g^{(2)}_{12}(\tau) \approx g^{(2)}_{2}(\tau)$ or leaving all parameters for $g^{(2)}_{12}(\tau)$ free in the fits for $g^{(2)}_{2}(\tau)$. The single-photon purity for the auto- and cross-correlations are also invariant  whether we  assume that the spectral shape of ZPL$_2$ is the same after its spectral jump or whether we allow it to be modified as would be the case if ZPL$_2$'s large emission prior to the spectral jump was due to an uncorrelated emitter spectrally overlapped with ZPL$_2$; in this case, we left all spectral parameters free to estimate the lineshape for ZPL$_2$. Under these different cases, the coefficient of determination ranged between 0.88-0.89, substantiating the claim that these assumptions have no affect on the single-photon purity of each ZPL or the magnitude of anti-correlations between ZPL$_1$ and ZPL$_2$. 

\subsection{Two-Color HBT Interferometry Discussion}

The auto- and cross-correlation functions and spectra for each emitter presented in Figures 1 and 2 provide experimental evidence for multiple excited and shelving states within a single hBN defect or complex of defects as shown in Figure 2e. Boron interstitials are potential candidate defects since they have charge states within the range of ZPL$_1$ and ZPL$_2$. However, due to the very low migration energy barriers (0.5 to 1 eV)\cite{Weston2018} for the charged states of B$_i$ we expect B$_i$ to photobleach rapidly in these experiments or to form a complex. Instead we suggest the ZPL$_1$ and ZPL$_2$ emission may be due to a complex of oxygen based defects or emitters with a X$_B$V$_N$ geometry such as N$_B$V$_N$ and C$_B$V$_N$ defects.  Alternatively, it is possible that ZPL$_1$ and ZPL$_2$ are associated with separate emitters in close proximity to one another such that the electronic dipole-dipole interaction between emitters results in the anti-correlations shown in Figure 2d. While defects have been found to be within 50~nm of one another\cite{hayee2020revealing}  electronic dipole-dipole interaction strengths at this distance are a negligible $\sim$10$^{-8}$~eV. In the limiting case of a separation distance between emitters of $\sim$1~nm the interaction strength would be $\sim$3~meV and would at best weakly couple these emitters. However we do not observe life-time enhancement relative to uncorrelated single emitters in our sample\cite{feldman2019phonon} or in the literature\cite{tran2016robust,tran2016quantum} and therefore reject the dipole-dipole coupling hypothesis.

\section{Summary and Conclusions}

In summary, we have observed correlated laser-irradiation-dependent changes in the quantum efficiency for two zero-phonon lines that reach equilibrium before simultaneously quenching in ambient laboratory conditions. Contrasting this with the photostability of emitters under vacuum suggests the presence of a photochemical process with oxygen that modifies the defect or lattice locally consistent with previous work~\cite{shotan2016photoinduced}. These zero-phonon lines were confirmed to be antibunched and  anti-correlated with one another. These cross-correlations clearly show photochromism between the ZPL$_1$ and ZPL$_2$ transitions, consistent with a similar study of the charged and neutral NV centers in diamond~\cite{berthel2015photophysics}.  These results are therefore strong evidence that ZPL$_1$ and ZPL$_2$ are excited state transitions for the same defect or complex. This first observation of photochromism in hBN defects is an essential step toward improved understanding of the atomic origins of these defects.

  
   
 

\section{Methods}
The sample studied in this work was a multilayer hBN flake (3-5 layers in thickness) from Graphene Supermarket. The flake was subsequently annealed in a First Nano rapid thermal processor at a temperature of 850$^o$C in 1 Torr N$_2$ with a temperature increase and decrease of 5$^o$C/min. For all experiments, a room-temperature confocal microscope with a 405 nm diode-laser excitation and a 0.9 NA objective was used to collect $\mu$PL from defects in our sample. Laser-edge filters and dichroic mirrors were used to pass only $\mu$PL to our spectrometer or interferometer. For laser-irradiation-dependent spectroscopy the $\mu$PL was passed to a diffraction grating spectrometer with 2~meV resolution. For singles counts and two-color HBT interferometry the $\mu$PL was passed to a beamsplitter with filters F1 and F2 prior to the detectors corresponding to arm one and two of the HBT interferometer. The detectors were fiber coupled using multimode fiber and spatial mode filtering was adjusted using varying fibers diameters (50-100 $\mu$m). Coincidence counts were collected using a Hyrdaharp 400 and the single photon detectors used were Perkin Elmer SPCM-AQR. 

\section{Acknowledgments}
This research was sponsored by the U. S. Department of Energy, Office of Science, Basic Energy Sciences, Materials Sciences and Engineering Division. Initial experimental planning and design was supported by the Laboratory-Directed Research and Development Program of Oak Ridge National Laboratory, managed by UT-Battelle, LLC for the U.S. Department of Energy.  M.F. gratefully acknowledges support by the Department of Defense (DoD) through the National Defense Science \& Engineering Graduate Fellowship (NDSEG) and NSF award DMR-1747426. C.E.M gratefully acknowledges postdoctoral research support from the Intelligence Community Postdoctoral Research Fellowship Program at the Oak Ridge National Laboratory, administered by Oak Ridge Institute for Science and Education through an interagency agreement between the U.S. Department of Energy and the Office of the Director of National Intelligence. Rapid thermal processing and spectroscopy experiments were carried out at the Center for Nanophase Materials Sciences (CNMS), which is sponsored at ORNL by the Scientific User Facilities Division, Office of Basic Energy Sciences, U.S. Department of Energy. The authors thank Harrison Prosper for discussion on uncertainty estimation for the auto- and cross-correlation fits. 

See Supplement Material 1 for supporting content.

\bibliography{references}


\end{document}